\documentclass[automark,headsepline,11pt,oneside,listsleft,DIV12]{scrartcl}
\usepackage{scrpage2}
\clearscrheadfoot
\ohead[]{\pagemark}
\usepackage{caption2}
\setcaptionwidth{1.0 \textwidth}

\usepackage{graphicx}  %added by ak
\usepackage{amssymb,amsmath} 
\usepackage{booktabs,units} 
\title{Calibrating Car-Following Models using Trajectory Data: Methodological Study}

\author{Arne Kesting$^1$ and Martin Treiber$^2$\\[1ex]Institute for Transport \& Economics\\
Technische Universit\"at Dresden (Germany)}
\date{March 28, 2008}
\begin{document}

\sloppy
\maketitle
\begin{abstract}
The car-following behavior of individual drivers in real city traffic
is studied on the basis of (publicly available) trajectory datasets
recorded by a vehicle equipped with an radar sensor. By means of a
nonlinear optimization procedure based on a genetic algorithm, we
calibrate the Intelligent Driver Model and the Velocity
Difference Model by minimizing the deviations between the observed
driving dynamics and the simulated trajectory when following the same
leading vehicle. The reliability and robustness of the nonlinear fits
are assessed by applying different optimization criteria, i.e.,
different measures for the deviations between two trajectories. The
obtained errors are in the range between~11\% and~29\% which is
consistent with typical error ranges obtained in previous
studies. Additionally, we found that the calibrated parameter values
of the Velocity Difference Model strongly depend on the optimization
criterion, while the Intelligent Driver Model is more robust in this
respect. By applying an explicit delay to the model input, we
investigated the influence of a reaction time. Remarkably, we found a
negligible influence of the reaction time indicating that drivers
compensate for their reaction time by anticipation. Furthermore, the
parameter sets calibrated to a certain trajectory are applied to the
other trajectories allowing for model validation. The results indicate
that ``intra-driver variability'' rather than ``inter-driver
variability'' accounts for a large part of the calibration errors. The
results are used to suggest some criteria towards a benchmarking
of car-following models.
\end{abstract}
\footnotetext[1]{E-mail: {\tt kesting@vwi.tu-dresden.de}, URL: {\tt http://www.akesting.de}, Postal address: Andreas-Schubert-Stra{\ss}e 23, 
D-01062 Dresden, Germany}
\footnotetext[2]{E-mail: {\tt treiber@vwi.tu-dresden.de}, URL: {\tt http://www.traffic-simulation.de}}
%%%%%%%%%%%%%%%%%%%%%%%%%%%%%%%%%%%%%%%%%%%%%%%%%%%%%%%%%%%%%%%%%%%%%%
%
%%%%%%%%%%%%%%%%%%%%%%%%%%%%%%%%%%%%%%%%%%%%%%%%%%%%%%%%%%%%%%%%%%%%%%
\newpage
\section*{Introduction}
As microscopic traffic flow models are mainly used to describe
collective phenomena such as traffic breakdowns, traffic
instabilities, and the propagation of stop-and-go waves, these models
are traditionally calibrated with respect to macroscopic traffic data,
e.g., one-minute flow and velocity data collected by double-loop
detectors. Nowadays, as microscopic traffic data have become more and
more available, the problem of analyzing and comparing microscopic
traffic flow models with real microscopic data has raised some
interest in the
literature~\cite{Brockfeld-benchmark04,Ranjitkar-bench04,Punzo-bench05,Ossen-benchmark05,Hoogendoorn-TRB07}.

In this paper, we will consider three empirical trajectories of
different drivers that are publicly available and that have been
provided by the Robert Bosch GmbH~\cite{Clearingstelle-DLR}. The
datasets have been recorded in 1995 during an afternoon peak hour on a
fairly straight one-lane road in Stuttgart, Germany. A car equipped
with a radar sensor in front provides the relative speed and distance
to the car ahead. The duration of the measurements are $250\,$s,
$400\,$s and $300\,$s, respectively. All datasets show complex
situations of daily city traffic with several acceleration and
deceleration periods including standstills due to traffic lights.
Because of their high resolution and quality, these datasets have
already been considered in the literature
before~\cite{Fouladvand-BoschData-2005,Panwai-Calibration-2005,Hoogendoorn-TRB07}.

We apply two car-following models of similar complexity
(thus, with the same number of parameters), namely the Intelligent
Driver Model~\cite{Opus} and the Velocity Difference
Model~\cite{Jiang-vDiff01} to the empirical trajectories. By means
of a nonlinear optimization, we will determine the ``optimal'' model
parameters which fit the given data best. In contrast to the previous
studies, we consider three different error measures because the fit
errors alone do not provide a good basis for an evaluation of the
applied models. Furthermore, we will argue that it is sufficient (and superior) to
minimize the objective functions exclusively with respect to the
vehicle gaps and not with respect of speeds. We show that the
variation in the parameter values with respect to the different
measures is surprisingly high for the Velocity Difference Model while
the Intelligent Driver Model is more robust suggesting a new criterion
in the context of ``benchmarking'' microscopic traffic models.  As
second contribution, we show that an additional parameter -- namely
the reaction time which is widely considered to be an important part
of a car-following model -- does not improve the reproduction of the
empirical data. Since reaction times clearly exist, this result
suggests that drivers compensate for the human reaction time by
anticipation.

In the following section, the two car-following models under
investigation will be introduced. In a second section, the
methodological approach for the nonlinear optimization problem will be
described. In particular, three different objective functions will be
introduced. The central third section presents the results of this
study: (i) The optimal model parameters for different objective
functions and for each of the three datasets will be summarized and
the simulated and empirical trajectories will be compared
directly. (ii) An alternative view will be given by the flow-density
relations resulting from the microscopic gaps and velocities. (iii)
The systematic variation of one parameter while keeping the others
constant allows for a one-dimensional scan of the models' parameter
spaces. (iv) By considering a delay in the input data, an explicit
reaction time will be studied as an additional parameter. (v) The
models will be validated by applying the determined optimal parameter
sets to the other datasets. Finally, we will close with a discussion
of the factors influencing the calibration errors and an outlook for
further work.

%%%%%%%%%%%%%%%%%%%%%%%%%%%%%%%%%%%%%%%%%%%%%%%%%%%%
%%%%%%%%%%%%%%%%%%%%%%%%%%%%%%%%%%%%%%%%%%%%%%%%%%%%
%
\section*{Car-Following Models under Investigation}
Microscopic traffic models describe the motion of each
individual vehicle, i.e., they model the action such as accelerations
and decelerations of each driver as a response to the surrounding
traffic by means of an acceleration strategy towards a desired
velocity in the free-flow regime, a braking strategy for approaching
other vehicles or obstacles, and a car-driving strategy for
maintaining a safe distance when driving behind another
vehicle. Microscopic traffic models typically assume that human
drivers react to the stimulus from neighboring vehicles with the
dominant influence originating from the directly leading vehicle known
as as ``follow-the-leader'' or ``car-following'' approximation.

In the following, we consider two microscopic car-following models
which are formulated as ordinary differential equations and,
consequently, space and time are treated as continuous variables. This
model class is characterized by an acceleration function $\dot{v}:=\frac{\text{d}v}{\text{d}t}$
that depends on the actual velocity $v(t)$, the (net distance) gap
$s(t)$ and the velocity difference $\Delta v(t)$ to the leading
vehicle: 
\begin{equation}\label{eq:mic}
\dot{v} (s,v,\Delta v) = f\left(s, v, \Delta v\right).
\end{equation}
Notice that we define $\Delta v$ as approaching rate, i.e., positive
if the following vehicle is faster than the leading vehicle.

\subsection*{Intelligent Driver Model}
The Intelligent Driver Model (IDM)~\cite{Opus} is defined by the acceleration
function
\begin{equation}\label{eq:IDMaccel}
\dot{v}_\text{IDM}(s, v, \Delta v) = a \left[ 1 -\left( \frac{v}{v_0} \right)^4 -\left(
         \frac{s^*(v,\Delta v)} {s}
         \right)^2 \right].
\end{equation}
This expression combines the acceleration strategy
$\dot{v}_\text{free} (v)= a[1-(v/v_0)^4]$ towards a {\it desired
velocity} $v_0$ on a free road with the parameter $a$ for the {\it
maximum acceleration} with a braking strategy $\dot{v}_\text{brake}(s,
v, \Delta v) = -a(s^*/s)^2$ which is dominant if the current gap
$s(t)$ to the preceding vehicle becomes smaller than the desired
minimum gap
\begin{equation}\label{eq:sstar}
s^*(v, \Delta v) = s_0 + v T + \frac{v \Delta v } {2\sqrt{a b}}.
\end{equation}
The {\it minimum distance} $s_0$ in congested traffic is significant
for low velocities only. The dominating term of Eq.~\eqref{eq:sstar}
in stationary traffic is $vT$ which corresponds to following the
leading vehicle with a constant {\it desired (safety) time gap} $T$.  The last
term is only active in non-stationary traffic and implements an
``intelligent'' driving behavior including a braking strategy that, in
nearly all situations, limits braking decelerations to the {\it
comfortable deceleration} $b$. Note, however, that the IDM brakes
stronger than $b$ if the gap becomes too small. This braking strategy
makes the IDM collision-free. All IDM parameters $v_0$, $T$, $s_0$,
$a$ and $b$ are defined by positive values.

\subsection*{Velocity Difference Model}
Another popular car-following model is the Velocity Difference
Model (VDIFF)~\cite{Jiang-vDiff01} which is closely related to the
Optimal Velocity Model by Bando et al.~\cite{Bando}. The
acceleration function consists of a term proportional to a
gap-dependent ``optimal velocity'' $v_\text{opt}(s)$ and a term that takes
velocity differences $\Delta v$ as a linear stimulus into account:
\begin{equation}\label{eq:acc_vdiff}
\dot{v}_\text{VDIFF}(s, v, \Delta v )  = \frac{v_\text{opt}(s)-v}{\tau} - \lambda \Delta v.
\end{equation}
The parameter $\tau$ is the {\it relaxation time} which describes the
adaptation to a new velocity due to changes in $s$ and $v$. The {\it
sensitivity parameter} $\lambda$ considers the crucial influence of
$\Delta v$. The properties of the VDIFF are defined by the function
for the optimal velocity $v_\text{opt}(s)$. In the literature, the
following function is proposed:
\begin{equation}\label{eq:vopt}
v_\text{opt}(s)=\frac{v_0}{2}\left[
\tanh\left(\frac{s}{l_\text{int}}-{\beta}\right) - \tanh(-{\beta})
\right].
\end{equation}
The parameter $v_0$ defines the {\it desired velocity} under free
traffic conditions. The {\it ``interaction length''} $l_\text{int}$
determines the transition regime for the $s$-shaped
function~\eqref{eq:vopt} going from $v_\text{opt}(s=0)=0$ to
$v_\text{opt}\to v_0$ when the distance to the leading vehicles
becomes large. Finally, the {\it ``form factor''} $\beta$ defines
(together with $l_\text{int}$) the shape of the equilibrium
flow-density relation (also known as fundamental diagram) which will
be considered below. In contrast to the IDM, the VDIFF exhibits
collisions for some regimes of the parameter space.

%%%%%%%%%%%%%%%%%%%%%%%%%%%%%%%%%%%%%%%%%%%%%%%%%%%%
%%%%%%%%%%%%%%%%%%%%%%%%%%%%%%%%%%%%%%%%%%%%%%%%%%%%

\section*{Calibration Methodology}
Finding an optimal parameter set for a car-following model with a
nonlinear acceleration function such as~\eqref{eq:IDMaccel} and
\eqref{eq:acc_vdiff} corresponds to a {\it nonlinear optimization
problem} which has to be solved numerically. Before the optimization
algorithm will be presented, we describe the simulation set-up and the
considered objective functions.

\subsection*{Simulation Set-Up}
The Bosch trajectory data~\cite{Clearingstelle-DLR} contains
velocities of both the leading and the following (measuring)
vehicle. These data therefore allow for a direct comparison between
the measured driver behavior and trajectories simulated by a
car-following model with the leading vehicle serving as externally
controlled input. Initialized with the empirically given distance and
velocity differences, $v^\text{sim}(t=0) = v^\text{data}(0)$ and
$s^\text{sim}(t=0) = s^\text{data}(0)$, the microscopic model is used
to compute the acceleration and, from this, the trajectory of the
following car. The gap to the leading vehicle is then given by the
difference between the simulated trajectory $x^\text{sim}(t)$ (front
bumper) and the given position of the rear bumper of the leading
vehicle $x^\text{data}_\text{lead}(t)$:
\begin{equation}
s^\text{sim}(t) = x^\text{data}_\text{lead}(t) - x^\text{sim}(t).
\end{equation}
This can be directly compared to the gap $s^\text{data}(t)$ provided
by the Bosch data. In addition, the distance $s^\text{sim}(t)$ has to
be reset to the value in the dataset when the leading object changes
as a result of a lane change of one of the considered vehicles. For
example, the leading vehicle of the dataset~3 (cf.\ the result section
below) turning into another street at $t\approx \unit[144]{s}$ which leads to a
jump in the gap of the considered follower.

%%%%%%%%%%%%%%%%%%%%%%%%%%%%%%%%%%%%%%%%%%
\subsection*{Objective Functions}
The calibration process aims at minimizing the difference between the
measured driving behavior and the driving behavior simulated by the
car-following model under consideration. Basically, any quantity can
be used as error measure that is not fixed in the simulation, such as
the velocity, the velocity difference, or the gap. In the following, we
use the error in the gap $s(t)$ for conceptual reasons: When
optimizing with respect to $s$, the average velocity errors are
automatically reduced as well. This does not hold the other way round,
as the error in the distance may incrementally grow when optimizing
with respect to differences in the velocities $v^\text{sim}(t)$ and
$v^\text{data}_\text{follow}(t)$.

For the parameter optimization, we need an objective function as
quantitative measure of the error between the simulated and observed
trajectories. As the objective function has a direct impact on the
calibration result, we consider three different error measures.  The
{\it relative error} is defined as a functional of the empirical and
simulated time series, $s^\text{data}(t)$ and $s^\text{sim}(t)$:
\begin{equation}\label{eq:rel_error_measure}
 \mathcal{F}_\text{rel}[s^\text{sim}] = \sqrt{ \left\langle \left(\frac{s^\text{sim} -
 s^\text{data}}{s^\text{data}} \right)^2 \right\rangle}\,.
\end{equation}
Here, the expression $\langle \cdot \rangle$ means the temporal average of a
time series of duration $\Delta T$, i.e.,
\begin{equation}
\langle z \rangle := \frac{1}{\Delta T} \int_{0}^{\Delta T} z(t)\,dt.
\end{equation}
Since the relative error is weighted by the inverse distance, this
measure is more sensitive to small distances $s$ than to large
distances.  As example, a simulated gap of \unit[10]{m} compared to a
distance of \unit[5]{m} in the empirical data results in a large error
of 100\%, whereas the same deviation of \unit[5]{m} leads, for
instance, to an error of 5\% only for a spacing of \unit[100]{m} which
is typical for large velocities.

In addition, we define the {\it absolute error} as
\begin{equation}\label{eq:abs_error_measure}
 \mathcal{F}_\text{abs}[s^\text{sim}] = \sqrt{ \frac{\left\langle(s^\text{sim}
 -s^\text{data} )^2 \right\rangle}{\left\langle s^\text{data} \right\rangle^2
 }}\,.
\end{equation}
As the denominator is averaged over the whole time series interval,
the absolute error $\mathcal{F}_\text{abs}[s^\text{sim}]$ is less
sensitive to small deviations from the empirical data than
$\mathcal{F}_\text{rel}[s^\text{sim}]$. However, the absolute error
measure is more sensitive to large differences in the numerator, i.e.,
for large distances $s$. Note that the error measures are normalized
in order to make them independent of the duration $\Delta T$ of the
considered time series allowing for a direct comparison of different
datasets.

As the absolute error systematically overestimates errors for large
gaps (at high velocities) while the relative error systematically
overestimates deviations of the observed headway in the low velocity
range, we will also study a combination of both error measures. For
this, we define the {\it mixed error measure}
\begin{equation}\label{eq:mix_error_measure}
 \mathcal{F}_\text{mix}[s^\text{sim}] = \sqrt{\frac{1}{\langle|s^\text{data}|\rangle} \left\langle
 \frac{(s^\text{sim} -
 s^\text{data})^2}{|s^\text{data}|}\right\rangle}\,.
\end{equation}

%%%%%%%%%%%%%%%%%%%%%%%%%%%%%%%%%%%%%%%%%%%%%%%%%%%%%%%%%%%
\subsection*{Optimization with a Genetic Algorithm}
For finding an approximative solution to the nonlinear optimization
problem, we will apply a genetic algorithm as search
heuristic~\cite{Goldberg-89}. The implemented genetic algorithm
proceeds as follows: (i) An ``individual'' represents a parameter set of
a car-following model and a ``population'' consists of $N$ such
sets. (ii) In each generation, the {\it fitness} of each individual in
the population is determined via one of the objective
functions~\eqref{eq:rel_error_measure},
\eqref{eq:abs_error_measure} or~\eqref{eq:mix_error_measure}.  (iii)
Pairs of two individuals are stochastically selected from the current
population based on their fitness score and recombined to generate a
new individual. Except for the best individual which is kept without
any modification to the next generation, the ``genes'' of all
individuals, i.e., their model parameters, are varied randomly
corresponding to a mutation that is controlled by a given
probability. The resulting new generation is then used in the next
iteration. (iv) The termination criterion is implemented as a two-step
process: Initially, a fixed number of generations is evaluated. Then,
the evolution terminates after convergence which is specified by a
constant best-of-generation score for at least a given number of
generations.

\subsection*{Parameter Constraints and Collision Penalty}
Both the IDM and the VDIFF contain $5$ parameters and are therefore
formally equivalent in their complexity.  In order to restrict the
parameter space for the optimization to reasonable and positive
parameter values without excluding possible solutions, we apply the
following constraints for the minimum and maximum values. For the IDM,
the desired velocity $v_0$ is restricted to the interval
$[1,70]\,\text{m/s}$, the desired (safety) time gap $T$ to
$[0.1,5]\,\text{s}$, the minimum distance $s_0$ to
$[0.1,8]\,\text{m}$, the maximum acceleration $a$ and the comfortable
deceleration $b$ to $[0.1,6]\,\text{m/s}^2$. For the VDIFF, the
allowed parameter intervals are $[1,70]\,\text{m/s}$ for the desired
velocity $v_0$, $[0.05, 20]\,\text{s}$ for the relaxation time $\tau$,
$[0.1, 100]\,\text{m}$ for the interaction length $l_\text{int}$ and
$[0.1, 10]\,\text{m}$ for the form factor $\beta$, and the (unit-less)
{\it sensitivity parameter} $\lambda$ is limited to $[0,3]$.

Last but not least, we have to take into account that some regions of
the VDIFF parameter space lead to collisions. In order to make these
``solutions'' unattractive to the optimization algorithm, we have
added a large crash penalty value to the objective measure which is
the standard procedure for numerical optimization.

%%%%%%%%%%%%%%%%%%%%%%%%%%%%%%%%%%%%%%%%%%%%%%%%%%%%
%%%%%%%%%%%%%%%%%%%%%%%%%%%%%%%%%%%%%%%%%%%%%%%%%%%%
%
\section*{Calibration Results}
\subsection*{Optimal Model Parameters} 
\begin{figure}
\centering
\begin{tabular}{cc}
\includegraphics[width=0.45\linewidth]{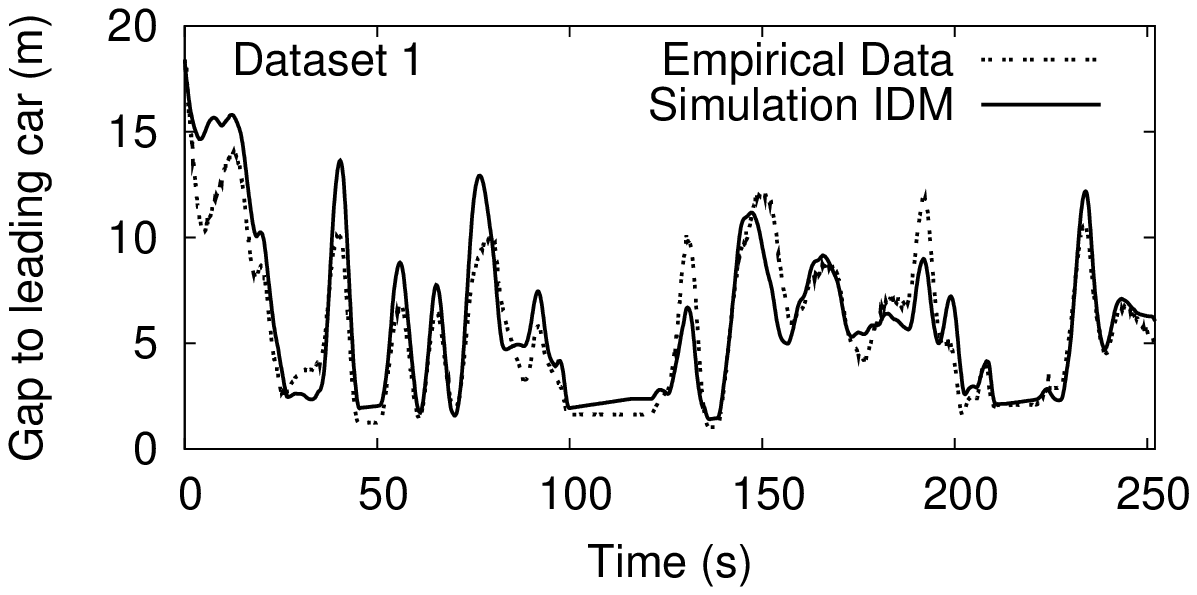} &
\includegraphics[width=0.45\linewidth]{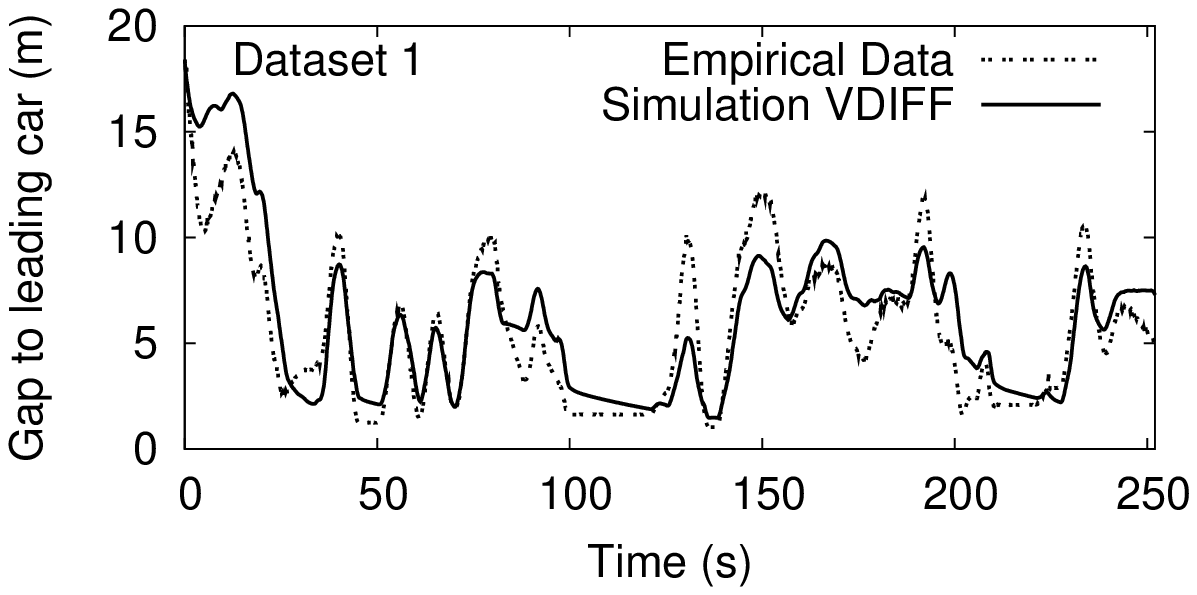} \\[-2mm]
\includegraphics[width=0.45\linewidth]{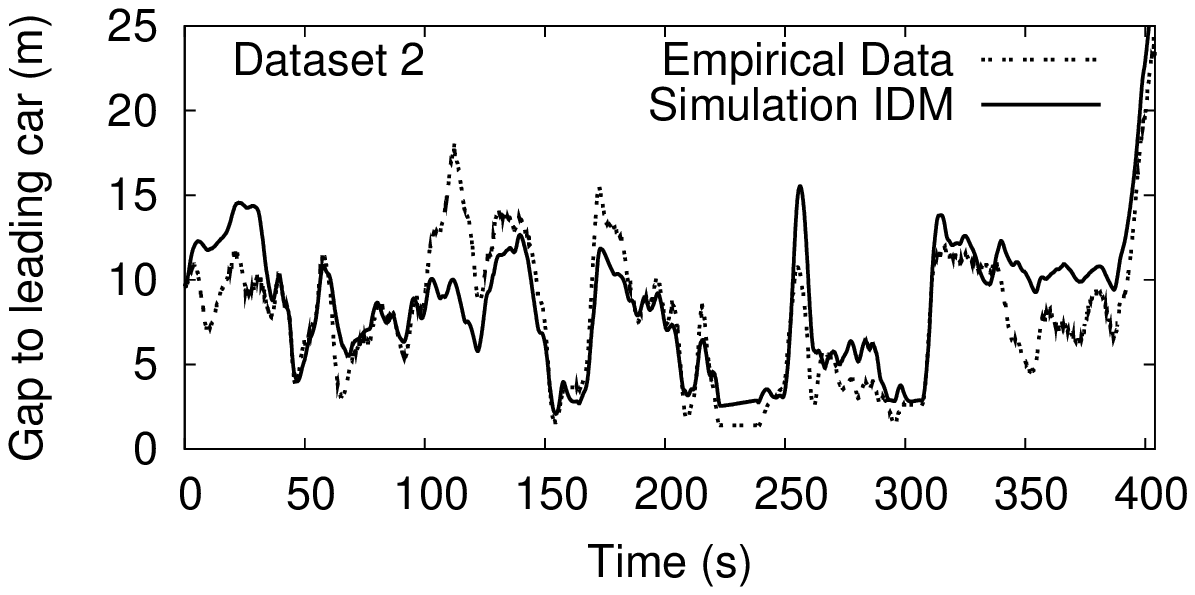} &
\includegraphics[width=0.45\linewidth]{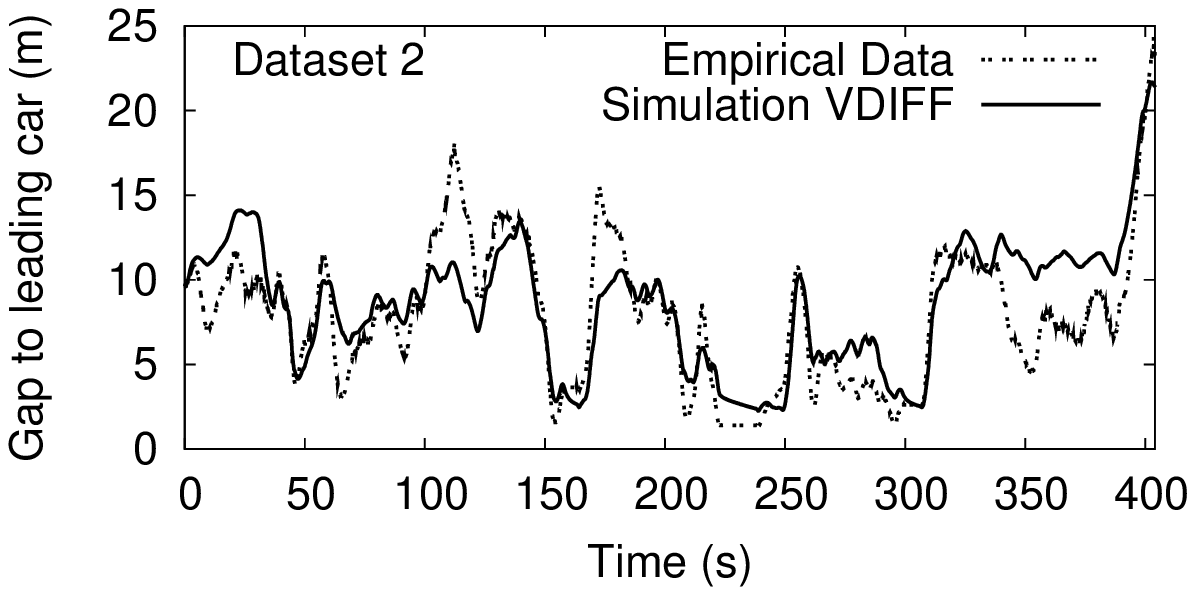} \\[-2mm]
\includegraphics[width=0.45\linewidth]{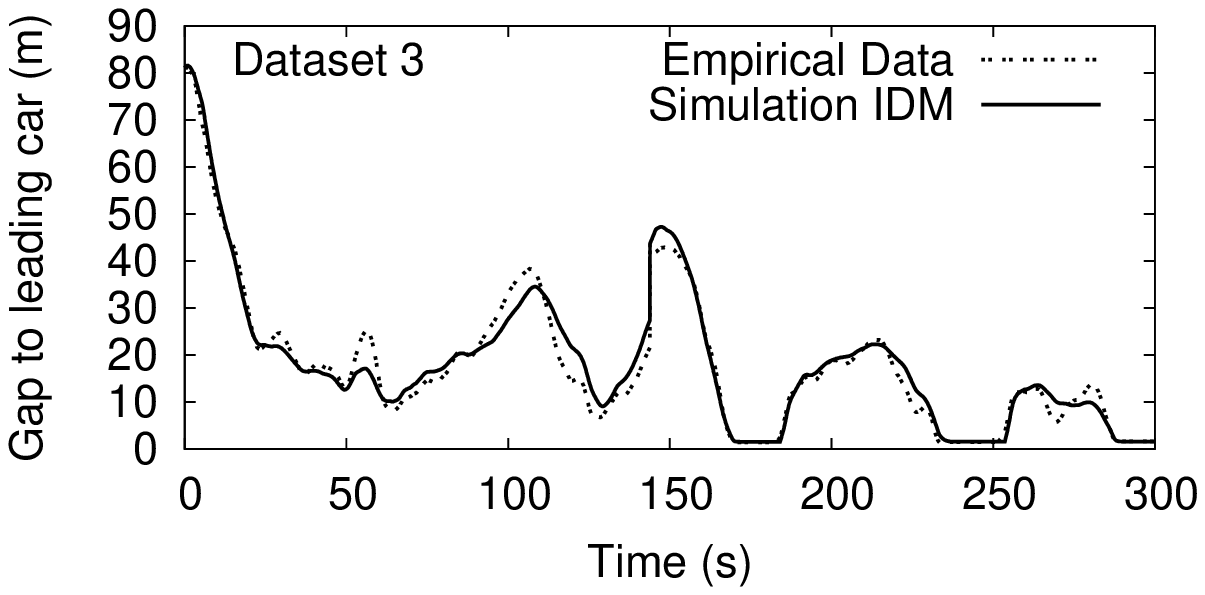} &
\includegraphics[width=0.45\linewidth]{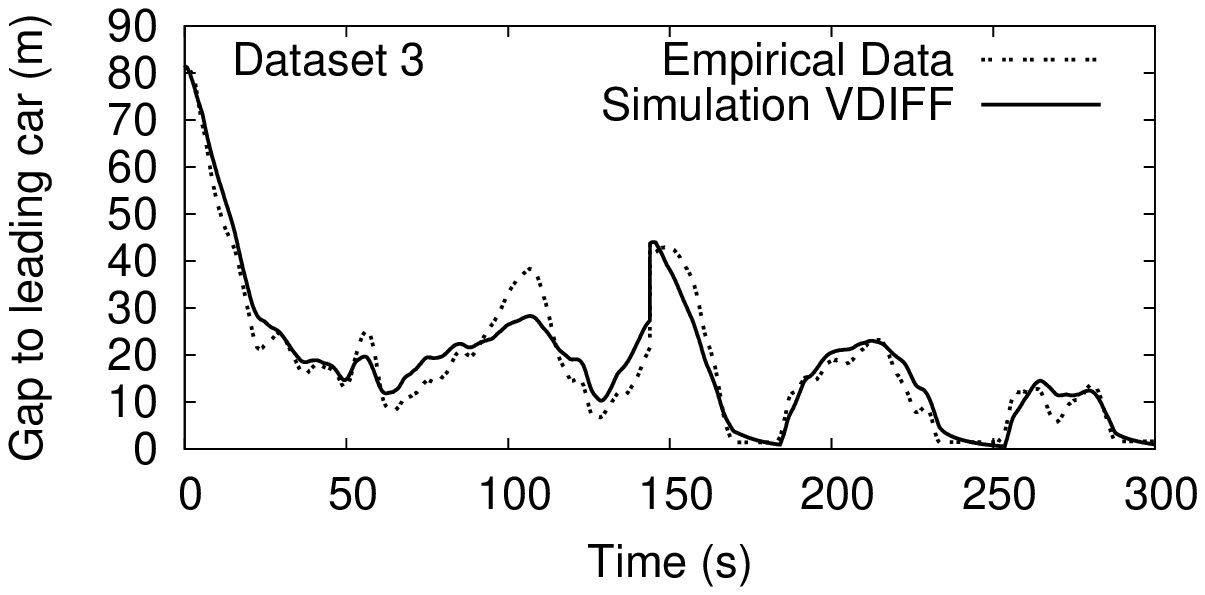}
\end{tabular}

 \caption{\label{fig:trajectories}Comparison of simulated and
 empirical trajectories. The model parameters are calibrated according
 to Table~\ref{tab:cal_param} for the mixed error
 measure~\eqref{eq:mix_error_measure}.}

\end{figure}

By applying the described optimization method, we have found the best
fit of the car-following models to the empirical data. The calibration
results for the three datasets and the considered three objective
functions~\eqref{eq:rel_error_measure},
\eqref{eq:abs_error_measure} and \eqref{eq:mix_error_measure} are
summarized in Table~\ref{tab:cal_param}. Furthermore,
Fig.~\ref{fig:trajectories} compares the dynamics of the gap $s(t)$
resulting from the calibrated parameters with the empirically measured
trajectories. The depicted simulations have been carried out with the
optimal parameters regarding the mixed error
measure~\eqref{eq:mix_error_measure}. The obtained errors are in the
range between~11\% and~29\% which is consistent with typical error
ranges obtained in previous
studies~\cite{Brockfeld-benchmark04,Ranjitkar-bench04,Punzo-bench05}. In
the concluding section, we will discuss the influencing factors for
the deviations between empirical and simulated car-following behavior.

Obviously, the calibrated model parameters vary from one dataset to
another because of different driving situations. Furthermore, a model
that fits best a certain driver not necessarily does so for a
different driver: In dataset~3, the IDM performs considerably better
than the VDIFF, while hardly any difference is found for
set~2. Moreover, the calibrated model parameters also depend
considerably on the underlying objective function. For example, the
dataset~3 can be reproduced best while the dataset~2 leads to the
largest deviations -- consistently for both the IDM and the VDIFF.
Here, the IDM parameters show a significantly smaller variation for a
considered dataset than the VDIFF. This finding is relevant for a
benchmarking of traffic models: It is not sufficient to consider only
the fit errors, but the quality of the traffic model is also
determined by the {\it consistency} and {\it robustness} of the
calibrated parameters. In a subsequent section, we will therefore
study the models' parameter spaces by means of a sensitivity and
validation analysis.

Let us discuss the values of the desired velocity obtained with the
IDM. In set~3, the desired speed is estimated to be
$v_0=\unit[58.0]{km/h}$ (corresponding to the maximum velocity reached
in the recorded driving situations) while the other two sets result in
$v_0\approx \unit[250]{km/h}$ (corresponding to the maximum value
allowed in the numerical optimization). This unreasonably high value
can be explained by the fact that the datasets~1 and~2 describe bound
traffic without acceleration periods to the desired speed. Therefore,
the calibration result of $v_0$ is only relevant for a {\it lower
bound}. This is plausible because the derived velocity does not
influence the driving dynamics if it is considerably higher than
$v_\text{lead}$ in a car-following situation. Consistent with this,
the error measures for the sets~1 and~2 hardly change when varying
$v_0$ in the range between \unit[60]{km/h} and \unit[250]{km/h}.

%%%%%%%%%%%%%%%%%%%%%%%%%%%%%%%%%%%%%%%%%%%%%%%%%%%%%%%%%%%%%%%%%%%%%%%%
\begin{table}[t]
{\small
\centering
\begin{tabular}{|l||r|r|r||r|r|r||r|r|r|}
\hline
{\bf IDM}  & \multicolumn{3}{c||}{Dataset 1} & \multicolumn{3}{c||}{Dataset 2} & \multicolumn{3}{c|}{Dataset 3} \\\hline\hline
Measure &   $\mathcal{F}_\text{rel}[s]$ & $\mathcal{F}_\text{mix}[s]$ & $\mathcal{F}_\text{abs}[s]$ & $\mathcal{F}_\text{rel}[s]$ & $\mathcal{F}_\text{mix}[s]$ & $\mathcal{F}_\text{abs}[s]$ & $\mathcal{F}_\text{rel}[s]$ & $\mathcal{F}_\text{mix}[s]$ & $\mathcal{F}_\text{abs}[s]$ \\\hline
Error $[\%]$          & 24.0  & 20.7 &  20.7  & 28.7   & 26.2  &  25.6  & 18.0  & 13.0  & 11.2  \\\hline\hline
$v_0\, [\text{m/s}]$  &  70.0 &  69.9 &  70.0  &  69.8  &  69.9 &  69.9  & 16.1  & 16.1  & 16.4\\
$T\, [\text{s}]$      &  1.07 &  1.12 &  1.03  &  1.51  &  1.43 &  1.26  & 1.30  & 1.30  & 1.39\\
$s_0\, [\text{m}]$    &  2.41 &  2.33 &  2.56  &  2.63  &  2.82 &  3.40  & 1.61  & 1.52  & 1.04\\
$a\, [\text{m/s}^2]$  &  1.00 &  1.23 &  1.40  &  0.956 &  0.977 &  1.06  & 1.58  & 1.56  & 1.52\\
$b\, [\text{m/s}^2]$  &  3.21 &  3.20 &  3.73  &  0.910 &  0.994 &  1.11  & 0.756 & 0.633 & 0.614\\\hline

\end{tabular}

\vspace{5mm}

\begin{tabular}{|l||r|r|r||r|r|r||r|r|r|}
\hline
\textbf{VDIFF}   & \multicolumn{3}{c|}{Dataset 1} & \multicolumn{3}{c|}{Dataset 2} & \multicolumn{3}{c|}{Dataset 3} \\\hline\hline
Measure & $\mathcal{F}_\text{rel}[s]$  & $\mathcal{F}_\text{mix}[s]$  & $\mathcal{F}_\text{abs}[s]$  & $\mathcal{F}_\text{rel}[s]$  & $\mathcal{F}_\text{mix}[s]$  & $\mathcal{F}_\text{abs}[s]$  & $\mathcal{F}_\text{rel}[s]$  & $\mathcal{F}_\text{mix}[s]$  & $\mathcal{F}_\text{abs}[s]$  \\\hline
Error [\%]                 & 25.5  & 25.8 & 21.4    & 29.1 & 26.7  & 25.6  & 28.2 & 19.0 & 14.5\\\hline\hline
$v_0 [\text{m/s}]$        & 7.02   & 14.8  & 18.1   & 11.7 & 49.5  & 9.56  & 70.0 & 26.3  & 46.2 \\
$\tau [\text{s}]$         & 11.9   & 20.0  & 4.90   & 1.48 & 20.0  & 20.0  & 19.4 & 4.87  & 5.45 \\
$l_\text{int} [\text{m}]$ & 1.62   & 9.60  & 5.23   & 3.93 & 12.1  & 4.26  & 28.6 & 20.7  & 40.9 \\
$\beta [\text{m}]$        & 4.16   & 1.21  & 2.14   & 2.69 & 1.89  & 2.30  & 1.31 & 0.758 & 0.102 \\
$\lambda [1]$             & 0.534  & 0.724 & 0.536  & 0.00 & 0.610 & 0.579 & 0.59 & 0.694 & 0.610 \\
\hline
\end{tabular}

  \caption{\label{tab:cal_param}Calibration results for the
  Intelligent Driver Model (IDM) and the Velocity Difference Model
  (VDIFF) for three different datasets and three different objective
  functions $\mathcal{F}$.}

}\end{table}

%%%%%%%%%%%%%%%%%%%%%%%%%%%%%%%%%%%%%%%%%%%%%%%%%%%%%%%%%%%

%%%%%%%%%%%%%%%%%%%%%%%%%%%%%%%%%%%%%%
%%%%%%%%%%%%%%%%%%%%%%%%%%%%%%%%%%%%%%

\subsection*{Microscopic Flow-Density Relations}
In the literature, the state of traffic is often formulated in
macroscopic quantities such as traffic flow and density. The
translation from the microscopic gap $s$ into the density $\rho$ is
given by the {\it micro-macro relation}
\begin{equation}\label{eq:micro_s}
\rho(s)  = \frac{1}{s+l},
\end{equation}
where $l$ is the vehicle length which we fix to $\unit[5]{m}$
here. The flow $Q$ as defined by the inverse of the time headway is
given by the vehicle's actual time gap $s/v$ and the passage time
for its own vehicle length $l/v$:
\begin{equation}
Q(s,v) = \frac{v}{s+l}.
\end{equation}

Furthermore, the flow-density points $(Q(t),\rho(t))$ can be
contrasted to the models' equilibrium properties describing states of
homogeneous and stationary traffic (so-called ``fundamental
diagrams'').  As equilibrium traffic is defined by vanishing velocity
differences and accelerations, the modeled drivers keep a constant
velocity $v_e$ which depends on the gap to the leading vehicle. For
the VDIFF, this equilibrium velocity is directly given by the optimal
velocity function~\eqref{eq:vopt}. Hence, the fundamental diagram
$Q(\rho)=v_e \rho$ can be directly calculated using
Eq.~\eqref{eq:micro_s}. For the IDM under the conditions $\dot{v}=0$ and
$\Delta v=0$ only the inverse, i.e., the equilibrium gap $s_e$ as
a function of the velocity, can be solved analytically leading to
\begin{equation}\label{eq:equil_idm}
s_{e}(v) = \frac{s_0+vT}{\sqrt{1 - \left(\frac{v}{v_0}\right)^{4}}}.
\end{equation}
However, the fundamental diagrams of the IDM can be
obtained numerically by parametric plots varying $v$.

\begin{figure}
\centering
\includegraphics[width=1.0\linewidth]{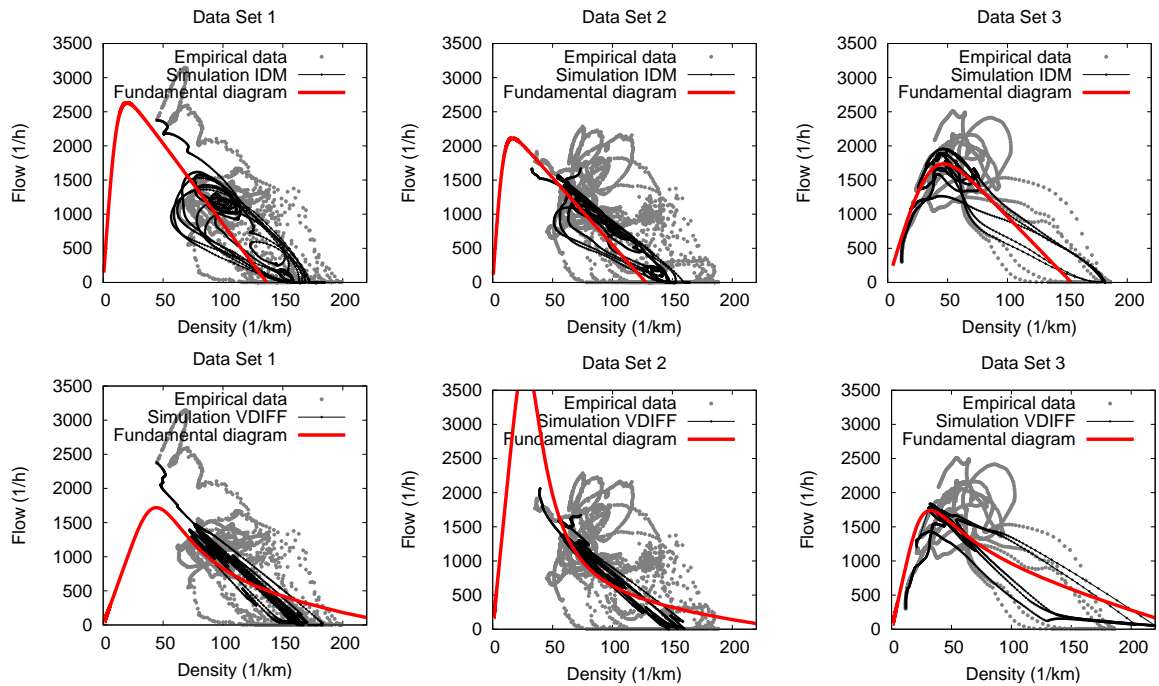}

 \caption{\label{fig:microFD}Microscopic flow-density relations
 $(Q(t), \rho(t))$ derived from the given and simulated gaps $s(t)$
 and velocities $v(t)$, respectively. In addition, the equilibrium
 flow-density relations (fundamental diagrams) are plotted as well. This
 representation offers an alternative view to
 Fig.~\ref{fig:trajectories} on the empirical and simulated data, in
 particular with respect to traffic stability.}

\end{figure}

In Fig.~\ref{fig:microFD}, the flow-density points $(Q(t),\rho(t))$
are plotted for each recorded time step of the empirical data and the
simulated trajectories. In addition, the fundamental diagram is
plotted as equilibrium curve. The diagrams give a good overview of the
recorded traffic situations. While sets~1 and~2 mainly contain
car-following behavior at distances smaller than $\unit[20]{m}$
(corresponding to densities larger than $\unit[50]{/km}$), the dataset
3 also features a non-restricted driving situation with a short period
of a free acceleration (corresponding to the branch with densities
lower than $\unit[30]{/km}$ of the flow-density plot). Furthermore,
the plots directly show the stability properties of the found optimal
parameter sets. Straight lines (for example in the datasets~1 and~2
for the VDIFF) correspond to very stable settings with short velocity
adaptation times $\tau$ while wide circles around the equilibrium
state (as for the IDM in set~1) indicate less stable settings
corresponding to smaller values of the IDM acceleration parameters
$a$~\cite{ThreeTimes-07}. Note that both parameters are related
inversely to each other: A large relaxation time $\tau$ in the VDIFF
corresponds to a small value of $a$ in the IDM.

%%%%%%%%%%%%%%%%%%%%%%%%%%%%%%%%%%%%%%
%%%%%%%%%%%%%%%%%%%%%%%%%%%%%%%%%%%%%%

\subsection*{Sensitivity Analysis}
Starting from the optimized model parameters summarized in
Table~\ref{tab:cal_param}, it is straightforward to vary a single
model parameter while keeping the other parameters constant. The
resulting one-dimensional scan of the parameter space gives a good
insight in the model's parameter properties and
sensitivity. Furthermore, the application of different
objective functions such as~\eqref{eq:rel_error_measure},
\eqref{eq:abs_error_measure} and \eqref{eq:mix_error_measure} can be seen as 
a benchmark for the {\it robustness} of the model calibration. A
``good'' model should not strongly depend on the chosen error measure.

\begin{figure}
\begin{center}
\begin{tabular}{cc}
\includegraphics[width=0.44\linewidth]{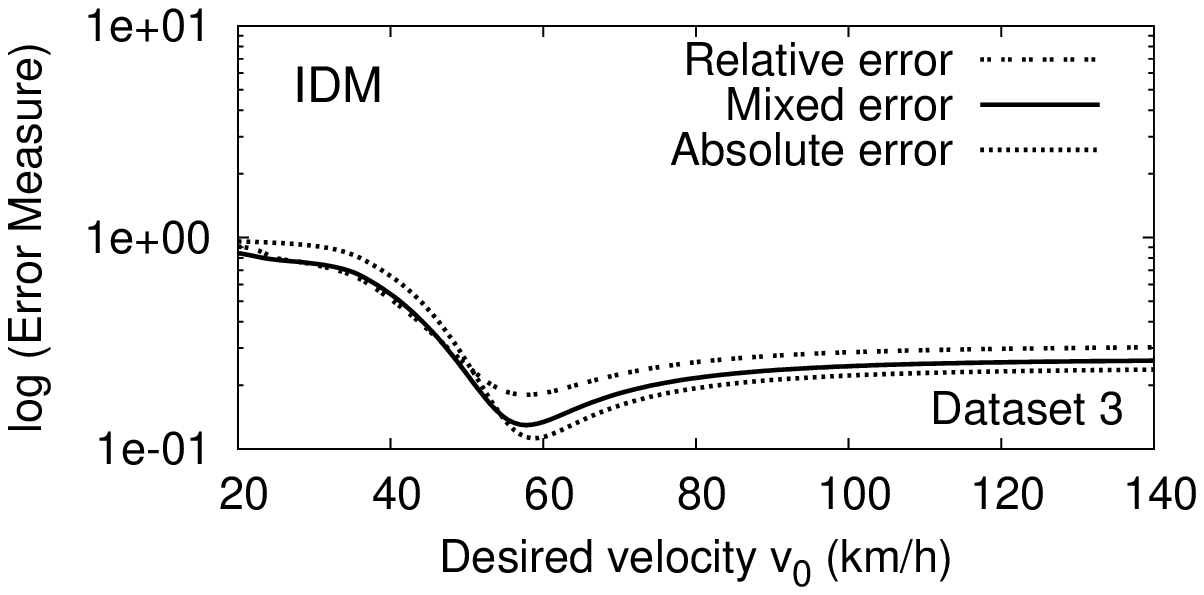} &
\includegraphics[width=0.44\linewidth]{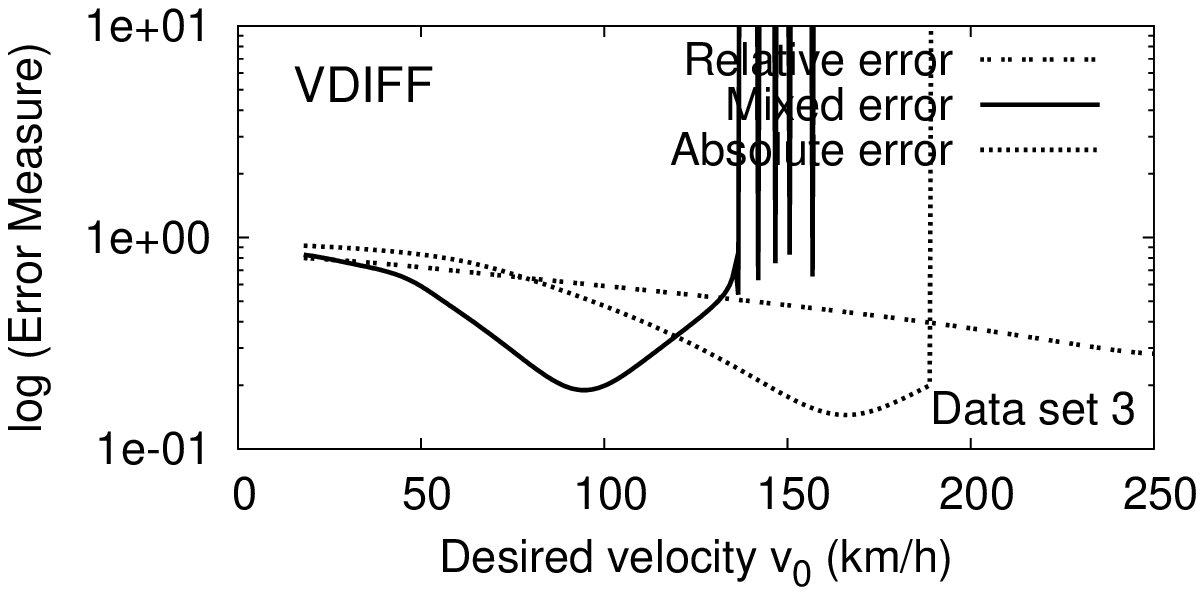}\\
\includegraphics[width=0.44\linewidth]{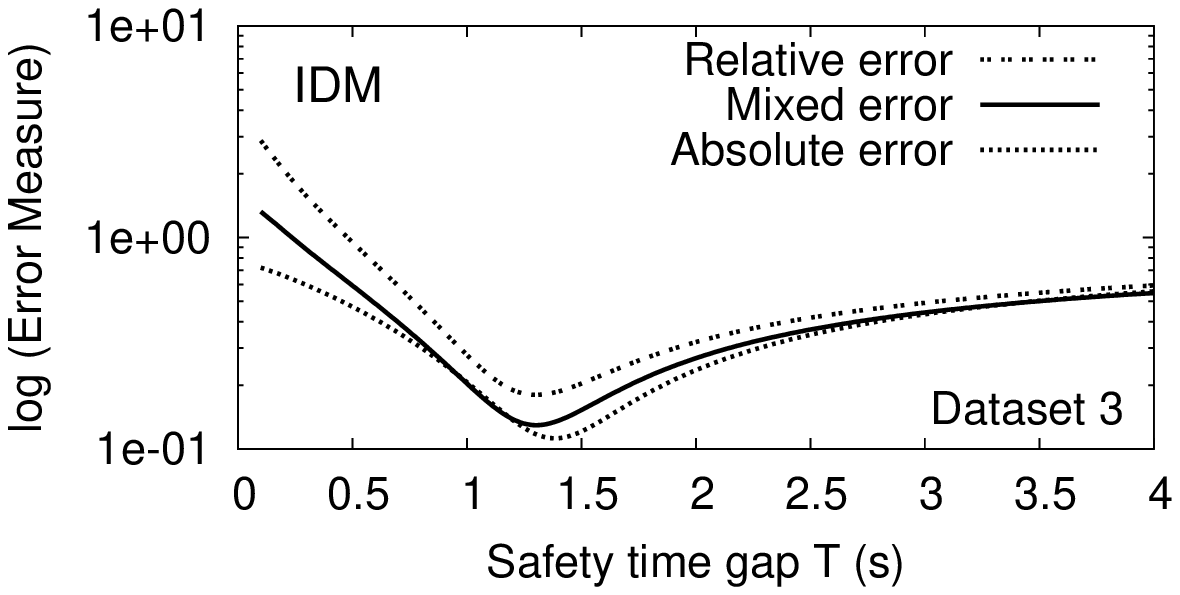} &
\includegraphics[width=0.44\linewidth]{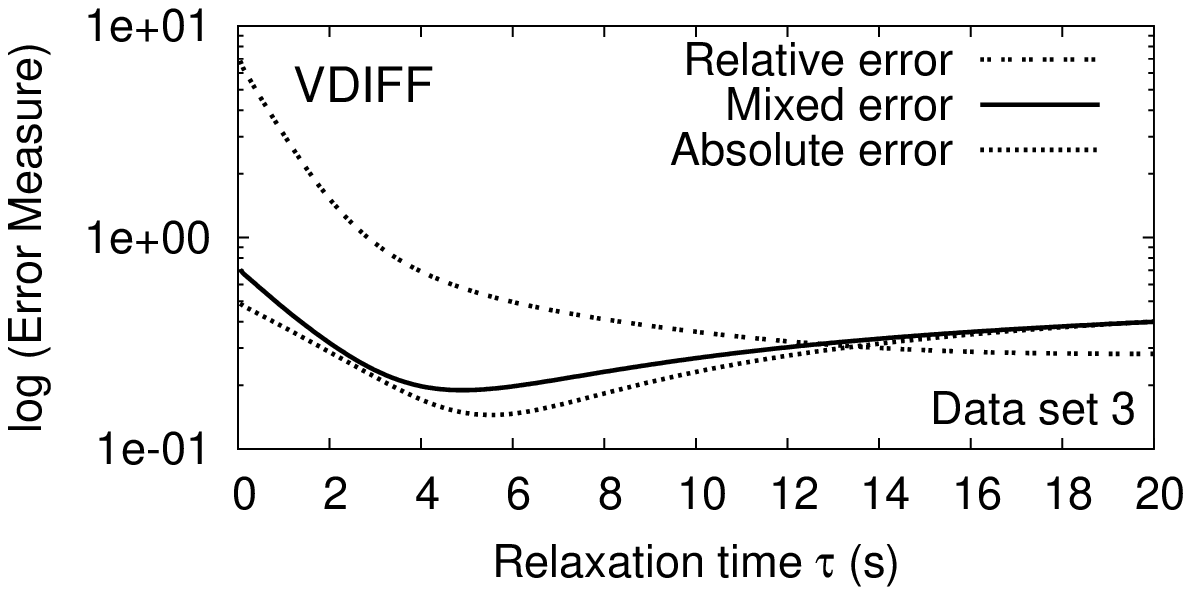} \\
\includegraphics[width=0.44\linewidth]{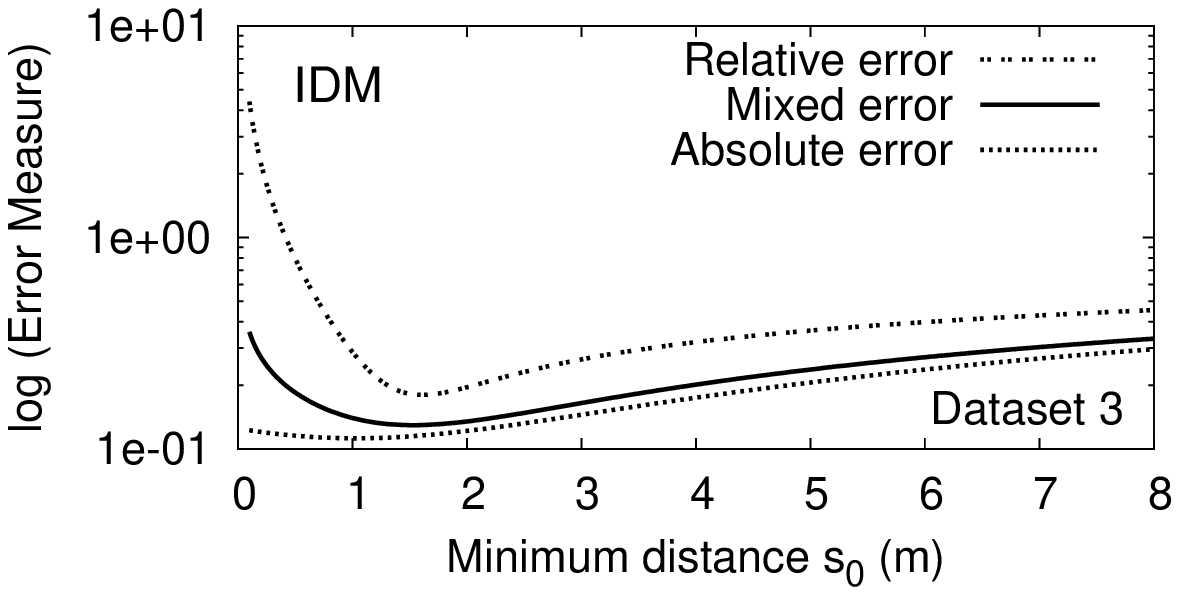} &
\includegraphics[width=0.44\linewidth]{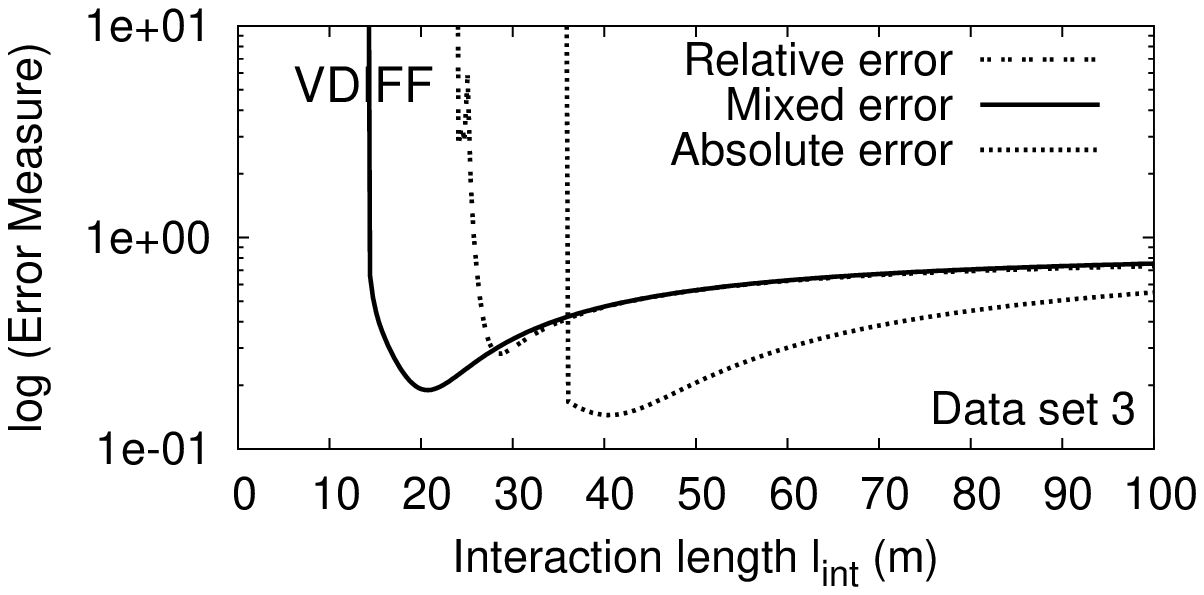} \\
\includegraphics[width=0.44\linewidth]{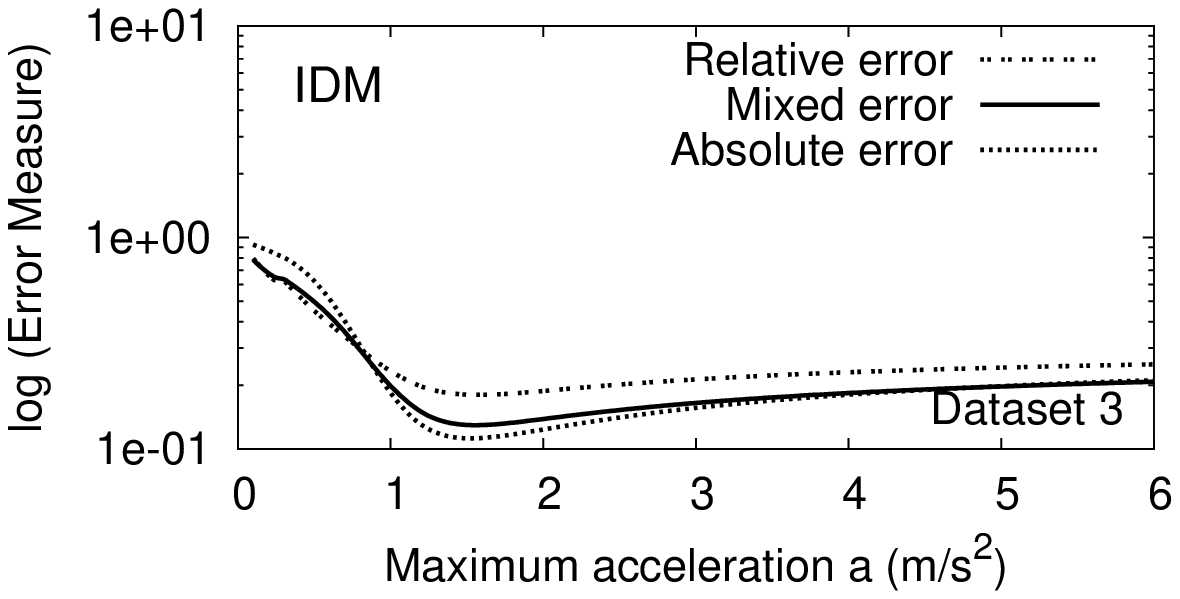} &
\includegraphics[width=0.44\linewidth]{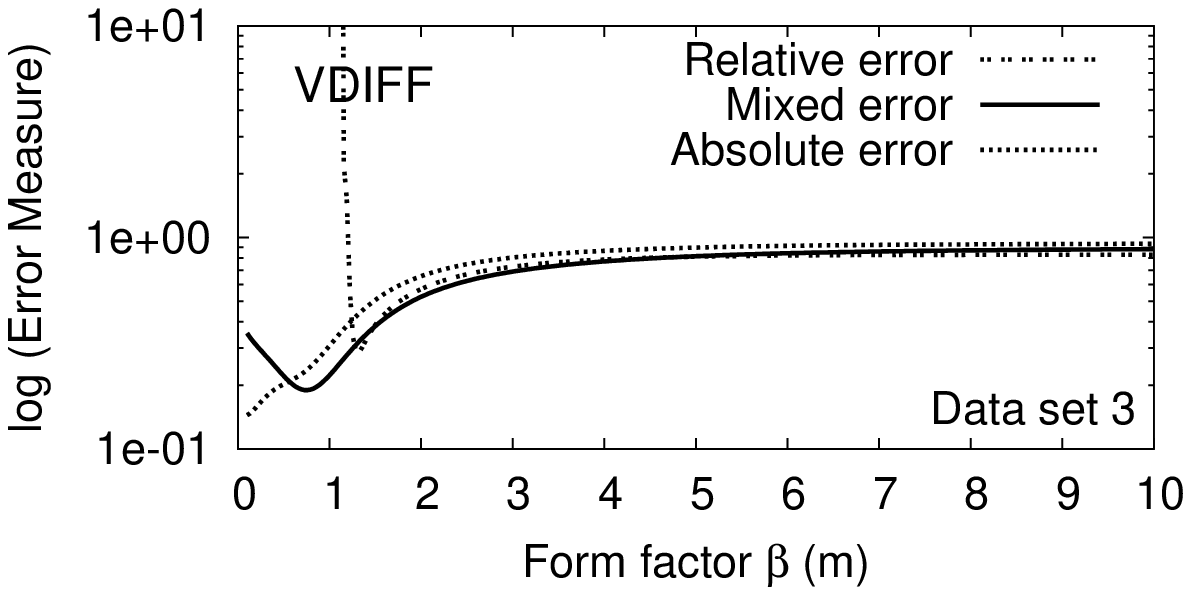} \\
\includegraphics[width=0.44\linewidth]{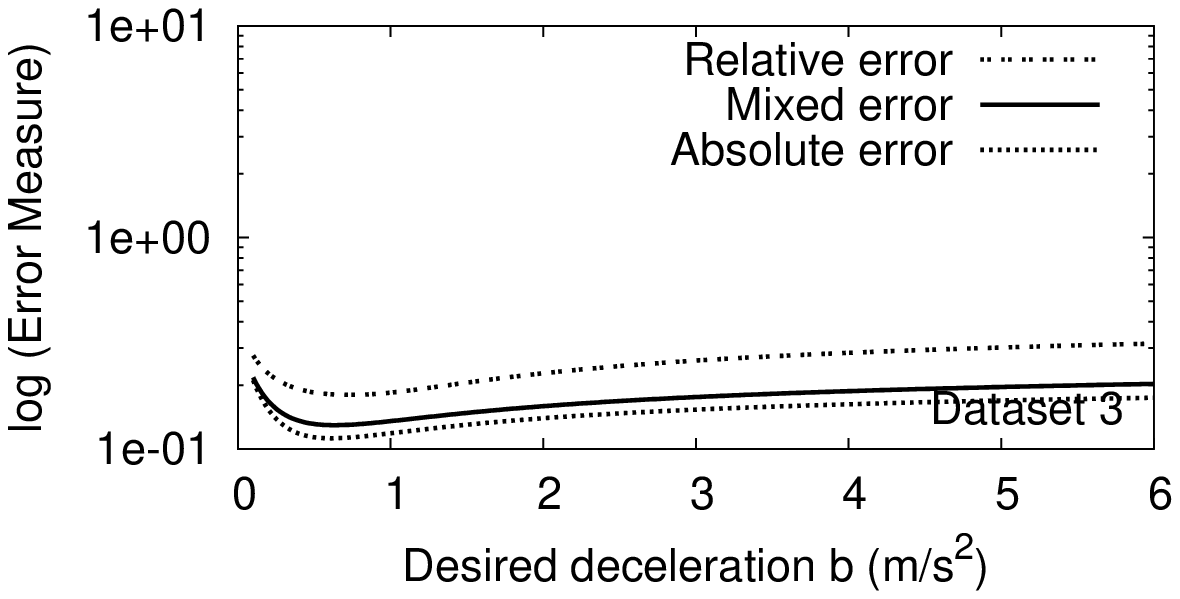} & 
\includegraphics[width=0.44\linewidth]{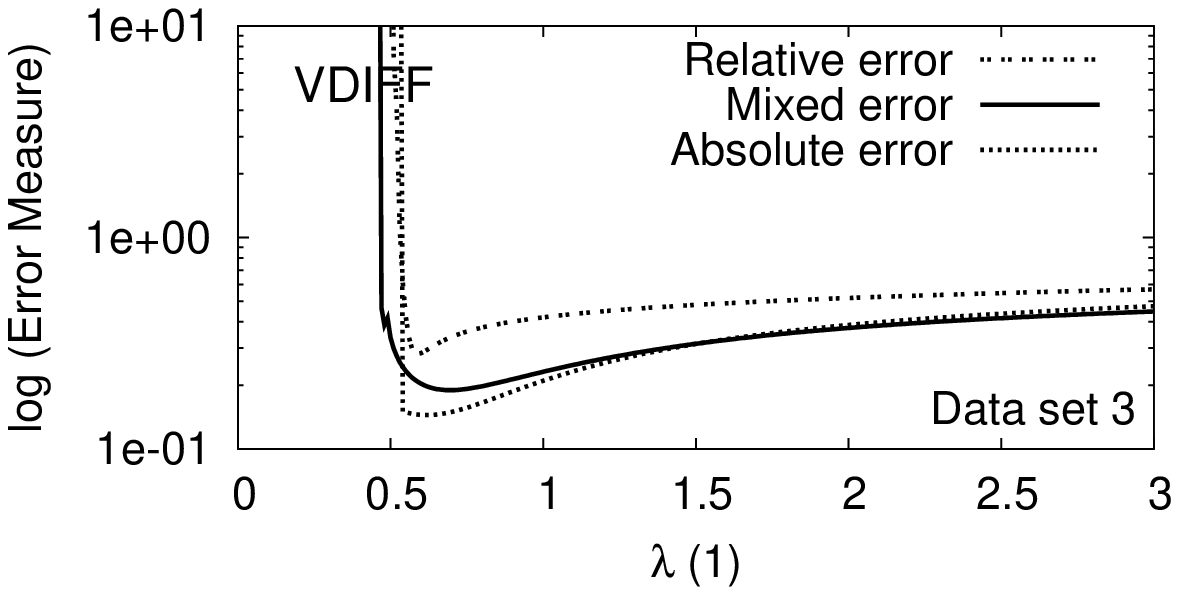}
\end{tabular}
\end{center}

 \caption{\label{fig:parameter_scans}Systematic variation of one model
 parameter while keeping the other parameters at the optimal values
 listed in Table~\protect\ref{tab:cal_param}. The diagrams show the
 considered error measures~\eqref{eq:rel_error_measure},
 \eqref{eq:abs_error_measure} and \eqref{eq:mix_error_measure} for the
 IDM (left column) and VDIFF (right column) using dataset~3. The
 errors are plotted in logarithmic scale. }
\end{figure}

Figure~\ref{fig:parameter_scans} shows the resulting error measures of
dataset~3. Remarkably, all error curves for the IDM are smooth and
show only one minimum (which is therefore easy to determine by the
optimization algorithm).  As the datasets mainly
describe car-following situations in obstructed traffic and
standstills, the IDM parameters $T$, $s_0$ and $a$ are particularly
significant and show distinct minima for the three proposed error
measures while the values of $v_0$ were hard to determine exactly from
the datasets~1 and~2 where the desired velocity is never
approximated. The comfortable deceleration $b$ is also not very
distinct (not shown here). The solutions belonging to different
objective functions are altogether in the same parameter range. This
robustness of the IDM parameter space is an important finding of this
study.

The results for the VDIFF imply a less positive model assessment: The
calibration results strongly vary with the chosen objective function
indicating a strong sensitivity of the model parameters. Furthermore,
too high values of the desired velocity lead to vehicle collisions in
the simulation as indicated by an abrupt raise in the error
curves. Interestingly, the sensitivity parameter $\lambda$ (taking
into account velocity differences) has to be larger than approximately
0.5 in order to avoid accidents. Velocity differences are therefore a
crucial input quantity for car-following models.

%%%%%%%%%%%%%%%%%%%%%%%%%%%%%%%%%%%%%%
%%%%%%%%%%%%%%%%%%%%%%%%%%%%%%%%%%%%%%

%%%%%%%%%%%%%%%%%%%%%%%%%%%%%%%%%%%%%%%%%%%%%%%%%%%%%%%%%%%%%%%%%%%%%%%%%%%%%
\subsection*{Consideration of an Explicit Reaction Time}
The considered car-following models describe an {\it instantaneous}
reaction (in the acceleration) to the leading car. A complex reaction
time is, however, an essential feature of human driving due to
physiological aspects of sensing, perceiving, deciding, and performing
an action. Therefore, it is interesting to incorporate a reaction time
in the IDM and the VDIFF and to investigate whether an additional
model parameter will improve the calibration results.

A reaction time $T_r$ can be additionally incorporated in a
time-continuous model of the type~\eqref{eq:mic} by evaluating the
right-hand side at a previous time $t-T_r$.  If the reaction time is a
multiple of the update time interval, $T_r=n
\Delta t$, it is straightforward to consider all input quantities at
$n$ time steps in the past. If $T_r$ is not a multiple of the update
time interval $\Delta t$, we use a {\it linear interpolation} proposed
in Ref.~\cite{HDM} according to
\begin{equation}\label{eq:intp}
x(t-T_r)=\beta x_{t-n-1}+(1-\beta) x_{t-n},
\end{equation}
where $x$ denotes any input quantity such as $s$, $v$ or $\Delta v$
(cf.\ the right-hand side of Eq.~\eqref{eq:mic}) and $x_{t-n}$ denotes
this quantity taken $n$ time steps before the actual step. Here, $n$
is the integer part of $T_r/\Delta t$, and the weight factor of the
linear interpolation is given by $\beta=T_r/\Delta t-n$.  As initial
conditions, values for the dependent variables are required for a
whole time interval $T_r$. In the simulations, we used as initial
conditions the values from the empirical data. For the stability
properties of the IDM with reaction time, we refer to
Ref.~\cite{ThreeTimes-07}.

\begin{figure}
\centering 
\begin{tabular}{c}
\includegraphics[width=0.45\linewidth]{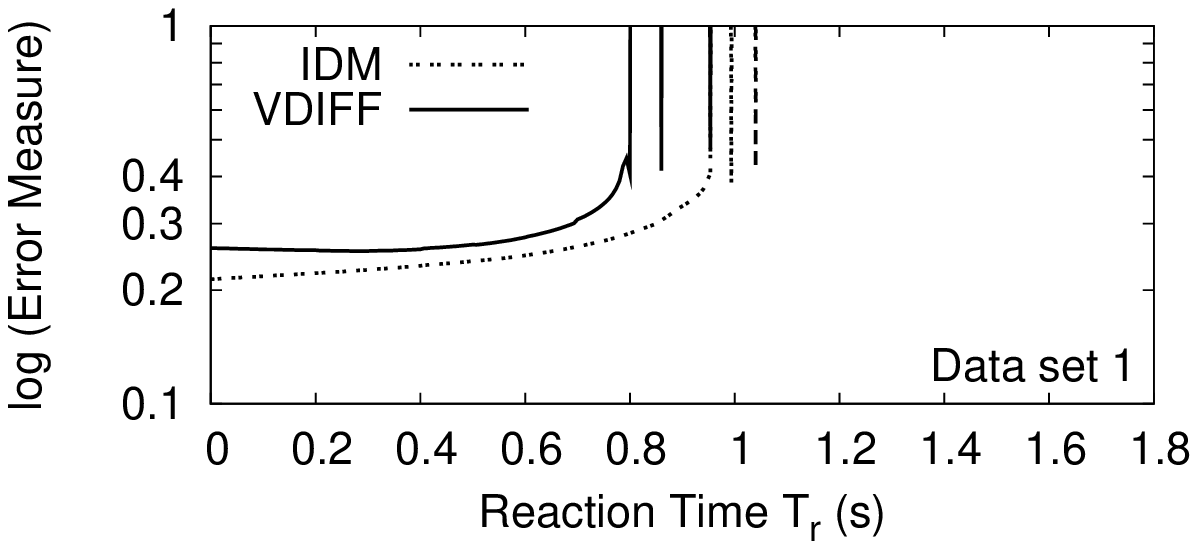} \\
\includegraphics[width=0.45\linewidth]{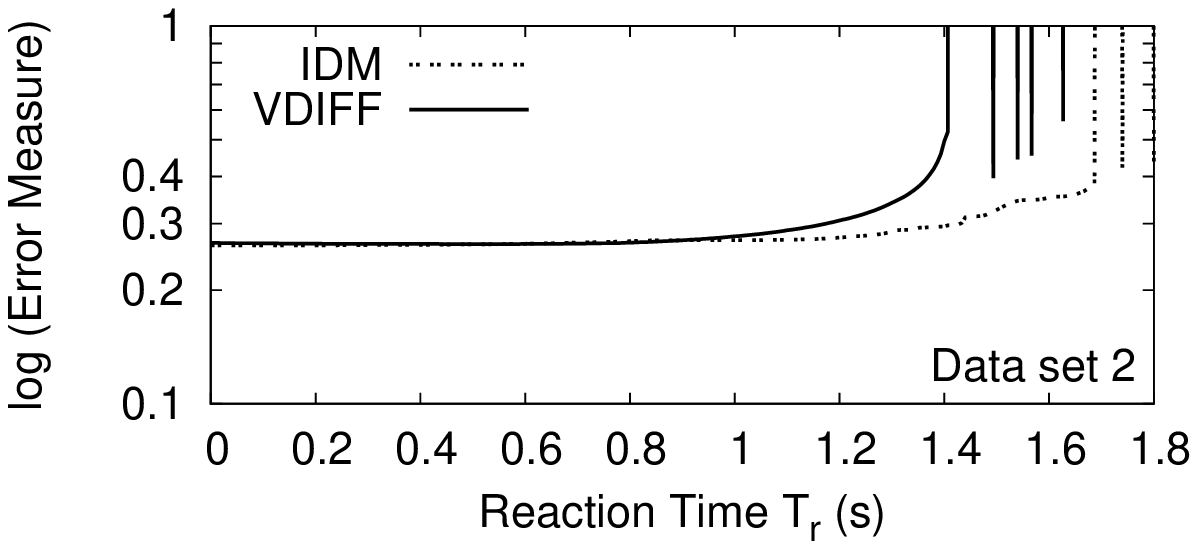} \\
\includegraphics[width=0.45\linewidth]{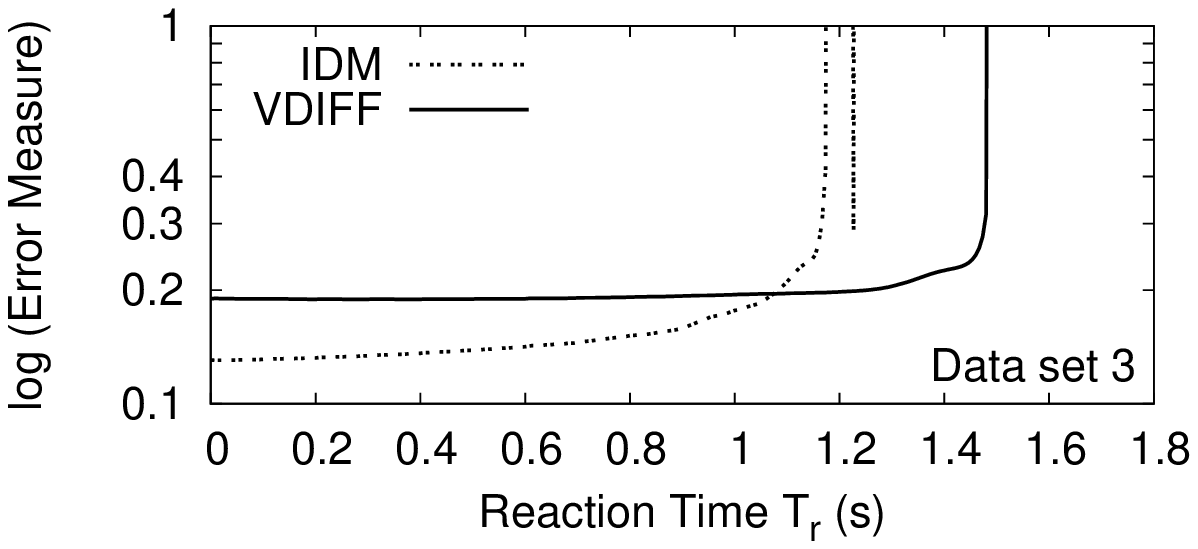} 
\end{tabular}

 \caption{\label{fig:scan_Tr}Systematic variation of a reaction time
 which has been explicitly incorporated in the Intelligent Driver
 Model (IDM) and the Velocity Difference Model (VDIFF). The reaction
 time $T_r$ has only a small influence for values well below the time
 gap $T$ which is an explicit model parameter of the IDM but only
 implicitly implemented in the VDIFF. Higher reaction times lead to
 collisions resulting in rapidly growing errors due to numerical
 penalties.}

\end{figure}

Figure~\ref{fig:scan_Tr} shows the systematic variation of the
reaction time $T_r$ while keeping the other parameters at their
optimal values as listed in Table~\ref{tab:cal_param} for the mixed
error measure~\eqref{eq:mix_error_measure}.  Interestingly, an
additional reaction time $T_r$ does {\it not} decrease the fit
errors. Moreover, for small reaction times, there is no influence at
all while values larger than a critical reaction time cause collisions
as indicated by the abrupt raise in the errors. For the IDM, this
critical reaction time $T_c^\text{crit}$ is smaller but of the order
of the calibrated time gap parameters for the three sets.  Similar
values have been found for the VDIFF. As the VDIFF does not feature an
explicit time gap parameter, however, it is not so easy to interpret.

The reason for the relatively high values of $T_c^\text{crit}$ is that
the considered scenarios are limited to a single pair of vehicles over
a limited duration and therefore only {\it local stability} properties
can be tested. This finding is in agreement with a simulation study on
local and collective stability properties of the IDM with explicit
delay~\cite{ThreeTimes-07,Treiber-TRR07}. Furthermore, the negligible
influence of the reaction time as an explanatory variable can be
interpreted in the way that the human drivers are able to compensate
for their considerable reaction time (which is about
\unit[1]{s}~\cite{green-reactionTimes}) by anticipation due to their
driving experience. These compensating influences have recently been
modeled and analyzed in~\cite{HDM,ThreeTimes-07}.

%%%%%%%%%%%%%%%%%%%%%%%%%%%%%%%%%%%%%%
%%%%%%%%%%%%%%%%%%%%%%%%%%%%%%%%%%%%%%

\subsection*{Validation by Cross-Comparison}

Let us finally validate the obtained calibrated parameters by applying
these settings to the other datasets, i.e., using the parameters
calibrated on the basis of another dataset. We use the three optimal
parameter settings listed in Table~\ref{tab:cal_param} and restrict
ourselves to the mixed error measure~\eqref{eq:mix_error_measure}.
The obtained errors can be found in
Table~\ref{tab:validation_param}. 

This {\it cross-comparison} allows to check for the reliability of the
obtained parameters and automatically takes into account the variance
of the calibrated parameter values. For the IDM, the obtained errors
for the cross-compared simulation runs are of the same order as for
the calibrated parameter sets. Therefore, the car-following behavior
of the IDM turned out to be robust with respect to reasonable changes
of parameter settings. In contrast, the VDIFF is more sensitive leading to larger
errors. One parameter set even led to collisions which is reflected in
a huge error due to the applied crash penalty.

%%%%%%%%%%%%%%%%%%%%%%%%%%%%%%%%%%%%%%%%%%%%%%%%%%%%%%%%%%%

\begin{table}[t!]
\centering

\begin{tabular}{lcccc}
\toprule
Model  & Dataset & Calib. Set 1 & Calib. Set 2 & Calib. Set 3 \\
\midrule
IDM & Set 1 & \underline{20.7\%} & 28.8\% & 28.7\% \\
 &  Set 2 & 35.2\% & \underline{26.2\%} & 40.1\% \\
 & Set 3 & 41.1\% & 27.0\% & \underline{13.0\%}\\
\midrule
VDIFF & Set 1 & \underline{25.8\%} & 64.3\% &  40.5\% \\
 & Set 2 &  28.8\% &  \underline{26.7\%} & 39.6\% \\
 & Set 3 &  57.0\% &  3840\% &  \underline{19.0\%}\%\\
%\midrule
%OVM & Set 1 & \underline{23.6\%} & 3760\% &  1500\% & 4300\%\\
%& Set 2 &  30.5\% &  \underline{26.7\%} & 48.9\% & 31.7\%\\
%&Set 3 &  2590\% &  561\% &  \underline{29.4\%} & 887\%\\
\bottomrule
\end{tabular}

 \caption{\label{tab:validation_param} Cross-comparison of the
 calibrated parameters for the mixed error
 measure~\eqref{eq:mix_error_measure} by applying the calibrated
 parameter sets for the Intelligent Driver Model (IDM) and the
 Velocity Difference Model (VDIFF) to the other datasets.  The
 underlined errors refer to the parameter values corresponding to the
 best calibration results.}

\end{table}

%%%%%%%%%%%%%%%%%%%%%%%%%%%%%%%%%%%%%%%%%%%%%%%%%%%%
%
%%%%%%%%%%%%%%%%%%%%%%%%%%%%%%%%%%%%%%%%%%%%%%%%%%%%
%
\section*{Discussion and Conclusions}
We have used the Intelligent Driver Model (IDM) and the Velocity
Difference Model (VDIFF) to reproduce three empirical trajectories. We
found that the calibration errors are between 11\% and 30\%. These
results are consistent with typical error ranges obtained in previous
studies~\cite{Brockfeld-benchmark04,Ranjitkar-bench04,Punzo-bench05}. Let
us finally discuss three qualitative influences which contribute to
these deviations between observation and reproduction. Note, however,
that noise in the data contribute to the fit errors as
well~\cite{Ossen-noise-TRR08}.

A significant part of the deviations between measured and simulated
trajectories can be attributed to the {\it inter-driver
variability}~\cite{Ossen-interDriver06} as it has been shown by
cross-comparison.  Notice that microscopic traffic models can easily
cope with this kind of heterogeneity because different parameter
values can be attributed to each individual driver-vehicle
unit. However, in order to obtain these distributions of calibrated
model parameters, more trajectories have to be analyzed, e.g., using
the NGSIM trajectory data~\cite{NGSIM}.

A second contribution to the overall calibration error results from a
non-constant driving style of human drivers which is also referred to
as {\it intra-driver variability}: Human drivers do not drive
constantly over time, i.e., their behavioral driving parameters
change. For a first estimation, we have compared the distances at
standstills in the dataset~3 with the minimum distance as direct model
parameter of the IDM. The driver stops three times because of red
traffic lights. The bumper-to-bumper distances are
$s_\text{stop,1}=\unit[1.39]{m}$, $s_\text{stop,2}=\unit[1.42]{m}$ and
$s_\text{stop,3}=\unit[1.64]{m}$. These different values in similar
situations already indicate that a deterministic car-following model
allows only for an averaged and, thus, ``effective'' description of
the human driving behavior resulting in parameter values that capture
the ``mean'' observed driving performance. Considering the theoretical
``best case'' of a perfect agreement between data and simulation for
all times except for the three standstills, the relative error
function depends on $s_0$ only and an analytical minimization of $s_0$
results in $s_{0}^\text{opt}\approx \unit[1.458]{m}$.  This optimal
solution defines a theoretical lower bound (based on about $15\%$ of
the data of the considered time series) for the relative error measure
of $\xi_\text{min}(s_0^\text{opt}) \approx 7.9\%$. Therefore, the
intra-driver variability accounts for a large part of the deviations
between simulations and empirical observations.  This influence could
be captured by considering {\it time-dependent} model parameters
reflecting driver adaptation processes as for example proposed
in~\cite{IDMM,VDT}.

Finally, {\it driver anticipation} contributes to the overall error as
well but is not incorporated in simple car-following models. This is
one possible cause for a {\it model error}, i.e., the residual
difference between a perfectly time-independent driving style and a
model calibrated to it.  For example, we found a negligible influence
of the additionally incorporated reaction time indicating that human
drivers very well anticipate while driving and therefore compensate
for their physiological reaction time. However, these physiological
and psychological aspects can only be determined indirectly by looking
at the resulting driving behavior. Consequently, it would be
interesting to check if a negative ``reaction'' (or rather
``anticipation'') time decreases the calibration errors. More complex
microscopic traffic models try to take those aspects into
account~\cite{HDM}.  Note, however, that multi-leader anticipation
requires trajectory data because the data recording using radar
sensors of single ``floating'' cars is limited to the immediate
predecessor~\cite{Hoogendoorn-multiAnti06}.

%%%%%%%%%%%%%%%%%%%%%%%%%%%%%%%%%%%%%%%%%%%%%%%%%%%%%%%%%%%

\textbf{Acknowledgment:}
We thank Dirk Helbing for a thorough review of the manuscript and
fruitful discussion.

%##########################################################
%References
%##########################################################
\bibliographystyle{osa}
%\bibliography{/home/vwitme/data/books/bibtex/bibtex/database,/home/vwitme/data/books/bibtex/bibtex/webcitations}

\end{document}